\documentclass[rnaas]{aastex62}

\usepackage{enumitem}
\usepackage{xspace,footmisc}
\usepackage{breakurl}

\newcommand{\json}{{\tt JSON}\xspace}
\newcommand{\csv}{{\tt CSV}\xspace}
\newcommand{\tsv}{{\tt TSV}\xspace}
\newcommand{\astroquery}{{\tt Astroquery}\xspace}
\newcommand{\astropy}{{\tt Astropy}\xspace}

\PassOptionsToPackage{linktocpage}{hyperref}

\begin{document}

\title{Open Astronomy Catalogs API}

%% Note that the corresponding author command and emails has to come
%% before everything else. Also place all the emails in the \email
%% command instead of using multiple \email calls.
\correspondingauthor{James Guillochon}
\email{jguillochon@cfa.harvard.edu}

\author[0000-0002-9809-8215]{James Guillochon}
\affiliation{Harvard-Smithsonian Center for Astrophysics\\
 60 Garden St.\\
 Cambridge, MA 02138, USA}

\author[0000-0002-2478-6939]{Philip S. Cowperthwaite}
\affiliation{Harvard-Smithsonian Center for Astrophysics\\
 60 Garden St.\\
 Cambridge, MA 02138, USA}

%% Note that RNAAS manuscripts DO NOT have abstracts.

%% Start the main body of the article. If no sections in the
%% research note leave the \section call blank to make the title.
\section{Description}

In this research note we announce the public release of the application program interface (API) for the Open Astronomy Catalogs \citep[OACs,][]{Guillochon:2017a}, the OACAPI\footnote{Source code available at \burl{https://github.com/astrocatalogs/OACAPI}}. The OACs serve near-complete collections of supernova\footnote{\burl{https://sne.space}}, tidal disruption\footnote{\burl{https://tde.space}} \citep{Auchettl:2017a}, kilonova\footnote{\burl{https://kilonova.space}} \citep{Villar:2017b}, and fast stars\footnote{\burl{https://faststars.space}} \citep{Boubert:2018a} data (including photometry, spectra, radio, and X-ray observations) via a user-friendly web interface that displays the data interactively and offers full data downloads. The API, by contrast, enables users to specifically download particular pieces of the OAC dataset via a flexible programmatic syntax.

While the description presented in this note is current with the active version of the API (1.0), a living version of this description is available on the {\tt git} repository for the API software\footnote{\burl{https://github.com/astrocatalogs/OACAPI}}, the reader is encouraged to bookmark that document to keep abreast of any future API changes. Note that {\it no API key is presently required for access}. Users are also able to run the API locally on their own computers and/or servers, in case a dedicated copy of the API is desired\footnote{A service setup for Linux is available at \burl{https://gist.github.com/guillochon/7148fe7a310cd4d08657f7e61c98dfe9}}.

\subsection{REST API}

The API is implemented via a Representational State Transfer (REST) service that allows users to access OAC data via {\tt GET} requests ({\tt POST} support coming soon), where the primary route to the data is in the form
\vspace{0.2em}
\begin{verbatim}
https://api.DOMAIN/OBJECT1+OBJECT2+.../QUANTITY1+QUANTITY2+.../
    ATTRIBUTE1+ATTRIBUTE2+...?ARGUMENT1=VALUE1&ARGUMENT2=VALUE2&...
\end{verbatim}
\vspace{0.2em}
where {\tt DOMAIN} corresponds to the catalog corresponding to the object type of the user's interest (e.g. {\tt sne.space} for supernovae, {\tt tde.space} for tidal disruptions, etc.), {\tt OBJECT} is set to a {\tt +}-delimited list of object names, {\tt QUANTITY} is set to a {\tt +}-delimited list of quantities to retrieve from all objects, {\tt ATTRIBUTE} is a {\tt +}-delimited list of properties of those quantities, and the {\tt ARGUMENT} variables (delimited by {\tt \&}) allow to user to filter data based upon various attribute values.

Alternatively, a user can locate objects by replacing the list of objects with {\tt catalog}
\vspace{0.2em}
\begin{verbatim}
https://api.DOMAIN/catalog/QUANTITY1+QUANTITY2+.../
    ATTRIBUTE1+ATTRIBUTE2+...?ARGUMENT1=VALUE1&ARGUMENT2=VALUE2&...
\end{verbatim}
\vspace{0.2em}
which will perform a search across the full catalog to retrieve the requested data. Routes that exclude the list of objects and/or quantities and/or attributes are also valid; the API will return {\it all} data corresponding to the full list of the omitted route components in these cases (see the examples in Table~\ref{tab:examples}).

Key names that are usable in API calls (for the {\tt QUANTITY} and {\tt ATTRIBUTE} parts of the API route) can be found in the OAC schema\footnote{\burl{https://github.com/astrocatalogs/schema}\label{foot:schema}}. The {\tt ARGUMENT} variables can be used to guarantee that a certain attribute appears in the returned results (e.g. adding {\tt \&time\&e\_magnitude} to the query will guarantee that each returned item has a time and {\tt e\_magnitude} attribute), and/or used to filter via a simple equality such as {\tt telescope=HST} (which would only return {\tt QUANTITY} objects where the telescope attribute equals ``HST''), and/or matched against regular expressions, and/or used for more sophisticated operations (an example being cone searches {\tt ra} and {\tt dec}). Below we show a list of special attributes that can be used to perform advanced queries:

\begin{description}[labelwidth=2cm,leftmargin =\dimexpr\labelwidth+\labelsep\relax, font=\sffamily\mdseries]
 \item[\tt closest] Return the quantities with the closest value to the specified attributes. If multiple attributes are specified, the closest to each will be returned (e.g., {\tt magnitude=15\&time=56789\&closest} would return both the observation with magnitude closest to 15 and time closest to 56789.
 \item[\tt complete] Return only quantities containing all of the requested attributes.
 \item[\tt first] Return only the first of each of the listed quantities.
 \item[\tt format=x] Return data in the specified format x, currently supports \csv and \tsv. Any other format specification will return \json.
 \item[\tt item=n] Return only the nth item of each of the listed quantities.
 \item[\tt radius=r] Return objects within a distance r (in arcseconds) of a given set of ra and dec coordinates. Note that this disables exact matches for ra and dec.
 \item[\tt width=w] Return objects within a distance w (in arcseconds) of a given ra value (for box searches).
 \item[\tt height=h] Return objects within a distance h (in arcseconds) of a given dec value (for box searches).
 \item[\tt sortby=s] Sort the returned array by the attribute {\tt s} (only works when returning results in \csv or \tsv formats).
\end{description}

In Table~\ref{tab:examples}, we provide some example queries that demonstrate the API's capabilities, these examples are hopefully useful to the reader as a starting point for OACAPI interactions.

\section{Astroquery module}

Along with the release of the REST API, we also announce the release of a module for the \astroquery package \citep[part of the \astropy package,][]{Astropy-Collaboration:2013a}, available in the {\tt 0.3.8} release of \astroquery\footnote{\burl{https://github.com/astropy/astroquery/releases/tag/v0.3.8}}. This module is designed to bring the full functionality of the REST API described above into a Pythonic workflow. The primary methods available to the user are:

\begin{description}[labelwidth=2cm,leftmargin =\dimexpr\labelwidth+\labelsep\relax, font=\sffamily\mdseries]
 \item[\tt query\_object] This method returns the requested {\tt QUANTITIES} and {\tt ATTRIBUTES} for a specified {\tt OBJECT} or list of {\tt OBJECTS}.
 \item[\tt query\_region] This method returns the results of a {\tt cone} or {\tt box} search for a single set of coordinates. Results can be filtered based on desired {\tt QUANTITIES} and {\tt ATTRIBUTES}.
\end{description}

Users can also quickly obtain results using the following tailored methods:

\begin{description}[labelwidth=2cm,leftmargin =\dimexpr\labelwidth+\labelsep\relax, font=\sffamily\mdseries]
 \item[\tt get\_photometry] This method returns all available photometry for a single {\tt OBJECT} or list of {\tt OBJECTS}.
 \item[\tt get\_single\_spectrum] This method returns a single spectrum for a single object at a specified {\tt MJD}.
 \item[\tt get\_spectra] This method returns all available spectra for a single {\tt OBJECT} or list of {\tt OBJECTS}.
\end{description}

All of the methods described here return an \astropy table constructed from the returned \csv query. Users can also request a \json compliant dictionary. We note that some searches, such as those returning multiple spectra, can not be processed into an \astropy table.

\begin{table*}
 \centering
 \begin{tabular}{p{8cm}|p{9cm}}
  User wants                                                                                                                             & URL route                                                                                                                                                                          \\
  \hline\hline
  All objects within a 2" cone about a set of coordinates                                                                                 & \burl{https://api.astrocats.space/catalog?ra=21:23:32.16                                                               & dec=-53:01:36.08  & radius=2}                              \\\hline
  Return all supernova metadata in \csv format                                                                                           & \burl{https://api.sne.space/catalog?format=csv}                                                                                                                                     \\\hline
  Redshifts of all supernovae with a redshift reported within $5^{\circ}$ of a coordinate, in \csv format                                & \burl{https://api.sne.space/catalog/redshift? ra=10:42:16.88                                                           & dec=-24:13:12.13  & radius=18000 & format=csv  & redshift} \\\hline
  Right ascensions and declinations of all objects with a redshift reported within $5^{\circ}$ of a coordinate, in \csv format            & \burl{https://api.astrocats.space/catalog/ra+dec?ra=10:42:16.88                                                        & dec=-24:13:12.13  & radius=18000 & format=csv  & redshift} \\\hline
  Get all references listed for an object in \csv format                                                                                  & \burl{https://api.sne.space/SN2014J/sources/reference+bibcode?format=csv}                                                                                                           \\\hline
  Select the first (preferred) value of the redshift                                                                                     & \burl{https://api.astrocats.space/SN2014J/redshift?first} or \burl{https://api.astrocats.space/SN2014J/redshift?item=0}                                                              \\\hline
  Return all photometric observations with at least one of the {\tt magnitude}, {\tt e\_magnitude}, and {\tt band} attributes            & \burl{https://api.astrocats.space/SN2014J/photometry/magnitude+e_magnitude+band}                                                                                                    \\\hline
  Return the above in \csv format                                                                                                        & \burl{https://api.astrocats.space/SN2014J/photometry/magnitude+e_magnitude+band?format=csv}                                                                                         \\\hline
  Only return observations that contain all requested attributes                                                                         & \burl{https://api.astrocats.space/SN2014J/photometry/magnitude+e_magnitude+band?complete}                                                                                           \\\hline
  Sort the returned photometry by magnitude in \csv format                                                                               & \burl{https://api.astrocats.space/SN2014J/photometry/magnitude+e_magnitude+band?format=csv                             & sortby=magnitude}                                          \\\hline
  Return observations for multiple objects at once, in \csv format                                                                        & \burl{https://api.astrocats.space/SN2014J+SN2015F/photometry/time+magnitude+band?format=csv}                                                                                        \\\hline
  Return only observations whose attributes include the listed keys ({\tt e\_magnitude} and {\tt band})                                  & \burl{https://api.astrocats.space/SN2014J/photometry/time+magnitude+e_magnitude+band?e_magnitude                       & band}                                                      \\\hline
  Return only observations matching given criteria, in this case {\tt band} = B                                                          & \burl{https://api.astrocats.space/SN2014J/photometry/magnitude+e_magnitude+band?band=B}                                                                                             \\\hline
  Luminosity distances and claimed types of all objects with an available luminosity distance and ``Ia'' listed as a type, in \tsv format & \burl{https://api.astrocats.space/catalog/lumdist+claimedtype?lumdist                                                  & claimedtype=ia    & format=tsv}                            \\\hline
  Claimed types of all objects with a typing that has a prefix matching ``Ia-'', in \tsv format &
\burl{https://api.astrocats.space/catalog/claimedtype?claimedtype=Ia-(.*)&format=tsv} \\\hline
  Return the spectrum closest to the listed MJD                                                                                          & \burl{https://api.astrocats.space/SN2014J/spectra/time+data?time=56703.2                                               & closest}                                                   \\\hline
  Return all photometry in a 2" radius about a coordinate, in \csv format                                                                & \burl{https://api.astrocats.space/catalog/photometry/time+band+magnitude?ra=21:23:32.16                                & dec=-53:01:36.08  & radius=2     & format=csv}             \\\hline
  Return the instruments used to produce spectra within a 5° of a given coordinate, in \csv format                                       & \burl{https://api.astrocats.space/catalog/spectra/instrument?ra=21:23:32.16                                            & dec=-53:01:36.08  & radius=18000 & format=csv}             \\\hline
 \end{tabular}
 \caption{Example API queries to perform various data retrieval tasks from the OACs. A full list of retrievable quantities is available in the \href{https://github.com/astrocatalogs/schema}{OAC schema}.}
 \label{tab:examples}
\end{table*}

\acknowledgments

We thank the regular users of the Open Astronomy Catalogs for testing the API as it was being deployed, with special thanks to Sebastian~Gomez for helpful comments. This work utilized the \astropy \citep{Astropy-Collaboration:2013a} and \astroquery \citep{Ginsburg:2018a} packages.

\bibliography{library}

\end{document}